\documentclass[12pt]{iopart}

\usepackage{graphicx}
\begin{document}

\title[Phase transitions of a tethered membrane model]{Phase transitions of a tethered membrane model with intrinsic curvature on spherical surfaces with point boundaries}

\author{Hiroshi Koibuchi}

\address{Department of Mechanical and Systems Engineering, Ibaraki National College of Technology, Nakane 866 Hitachinaka, Ibaraki 312-8508, Japan}
\ead{koibuchi@mech.ibaraki-ct.ac.jp}
\begin{abstract}
We found that the order for the crumpling transition of an intrinsic curvature model changes depending on the distance between two boundary vertices fixed on the surface of spherical topology. The model is a curvature one governed by an intrinsic curvature energy, which is defined on triangulated surfaces. It was already reported that the model undergoes a first-order crumpling transition without the boundary conditions on the surface. However, the dependence of the transition on such boundary condition is yet to be studied. We have studied in this paper this problem by using the Monte Carlo simulations on surfaces up to a size $N\!=\!8412$. The first-order transition changes to a second-order one if the distance increases.  

\end{abstract}

\maketitle

\section{Introduction}\label{intro}
Over the past two decades, a considerable number of studies have been performed on the phase structure of the surface model of Helfrich \cite{HELFRICH-1973}, Polyakov\cite{POLYAKOV-NPB1986}, and Kleinert \cite{KLEINERT-PLB1986} for biological membranes \cite{NELSON-SMMS2004,David-TDQGRS-1989,David-SMMS2004,Wiese-PTCP2000,Bowick-PREP2001,WHEATER-JP1994,Peliti-Leibler-PRL1985,DavidGuitter-EPL1988,PKN-PRL1988,BKS-PLA2000,BK-PRB2001}. The surface models can be classified into two groups, which are characterized by the curvature energy in the Hamiltonian; one is an extrinsic curvature model \cite{KANTOR-NELSON-PRA1987,KANTOR-SMMS2004,WHEATER-NPB1996,BCFTA-JP96-NPB9697,KY-IJMPC2000-2,KOIB-PLA-2003-2,KD-PRE2002,KOIB-PRE-2004-1,KOIB-PRE-2005-1,KOIB-NPB-2006,CATTERALL-NPBSUP1991,AMBJORN-NPB1993,ABGFHHM-PLB1993,BCHHM-NPB9393,KY-IJMPC2000-1,KOIB-PLA20023,KOIB-PLA-2004,KOIB-EPJB-2005}, and the other is an intrinsic curvature model \cite{BJ-PRD-1993-1994,BEJ-PLB-1993,BIJJ-PLB-1994,FW-PLB-1993}. The extrinsic curvature model is known to undergo a first-order transition between the smooth phase and the crumpled phase on tethered spherical surfaces \cite{KOIB-PRE-2004-1,KOIB-PRE-2005-1,KOIB-NPB-2006}. 

Studies have also focused on the phase structure of the model with intrinsic curvature \cite{KOIB-PRE-2003,KOIB-PRE-2004-2,KOIB-EPJB-2004,KOIB-PLA-2005-1,KOIB-PLA-2006-1}. It has been shown that the first-order transition can be seen in spherical fluid/tethered surfaces \cite{KOIB-PRE-2004-2,KOIB-EPJB-2004}, tethered surface of disk topology \cite{KOIB-PLA-2005-1}, and tethered surface with torus topology \cite{KOIB-PLA-2006-1}. 

However, little attention has been given to the boundary conditions of the surface models, although the center of the surface can be fixed to prevent translation in Monte Carlo (MC) simulations. Though, it is as yet unclear how boundary conditions of the surface influence the crumpling transition. 

Therefore, we should carefully study the influences of boundary conditions on the phase transition. In fact, we know that a phase transition is significantly influenced by boundary conditions, which fix some of the dynamical variables to a prescribed value. Moreover, it is clear that artificial vesicles can be supported with substrates such as glass plates or beads in an aqueous solution. 

In this paper, we study how a boundary condition influences the crumpling transition of the tethered surface model with intrinsic curvature on spherical surfaces. The boundary condition is imposed on the surface with two fixed vertices of distance $2L(N)$, which depends on the total number of vertices $N$ of the surface. We will find that the order of the transition changes from first-order to second-order and higher-orders in the limit of $N\!\to\!\infty$ when $L(N)$ is increased from $L_0(N)$ to $1.5L_0(N)$ and $2L_0(N)$, respectively, where $L_0(N)$ is a radius of the initial sphere such that the Gaussian energy $\langle S_1\rangle/N$ is approximately equal to $\langle S_1\rangle/N\!\sim\!3/2$ at the start of MC simulations. The result in this paper is in sharp contrast to that of the fluid surface model with extrinsic curvature, where the phase transition is strengthened with the increasing $L(N)$ \cite{KOIB-PLA-2004,KOIB-EPJB-2005}.

\section{Model and Monte Carlo technique}
The tethered surface model is defined by the partition function 
\begin{eqnarray} 
\label{Part-Func}
 Z = \int^\prime \prod _{i=1}^{N} d X_i \exp\left[-S(X)\right],\\  
 S(X)=S_1 + \alpha S_3, \nonumber
\end{eqnarray} 
where $\alpha$ is the curvature coefficient, and $\int^\prime$ denotes the boundary condition in which  two vertices are fixed and separated by a distance of $2L(N)$. $S(X)$ denotes that the Hamiltonian $S$ depends on the position variables $X$ of vertices. $S_1$ and $S_3$ are defined by
\begin{equation}
\label{Disc-Eneg} 
S_1=\sum_{(ij)} \left(X_i-X_j\right)^2,\quad S_3=-\sum_i\log( \delta_i/2\pi),
\end{equation} 
where $\sum_{(ij)}$ in $S_1$ is the sum over bonds $(ij)$ connecting the vertices $i$ and $j$. The bonds $(ij)$ are the edges of the triangles. $\delta_i$ in Eq. (\ref{Disc-Eneg}) is the vertex angle, which is the sum of the angles meeting at the vertex $i$. We call $S_3$ the deficit angle term, because $\delta_i\!-\!2\pi$ is just the deficit angle. We note that $\sum_i (\delta_i\!-\!2\pi)$ is constant on surfaces of fixed genus because of the Gauss-Bonnet theorem. 

We comment on the unit of physical quantities in the model. By letting $a$ be a length unit in the model, we can express all quantities with unit of length in terms of $a$. Hence, the unit of $S_1$ is $a^2$, and that of $L_0(N)$ is $a$. We fix the value of $a$ to $a\!=\!1$ in this paper, because the length unit can be arbitrarily chosen in the model. The unit of $\alpha$ is expressed by $kT$, where $k$ is the Boltzmann constant, $T$ is the temperature.  Note also that varying the temperature $T$ is effectively identical to varying $\alpha$ in the model.

Triangulated surfaces are obtained by dividing the icosahedron. By splitting the edges of the icosahedron into $\ell$-pieces, we have a surface of size $N\!=\!10\ell^2\!+\!2$. These surfaces are identical to those in \cite{KOIB-PRE-2005-1,KOIB-NPB-2006}. These surfaces are characterized by $N_5\!=\!12$ and $N_6\!=\!N\!-\!12$, where $N_q$ is the total number of vertices with a co-ordination number $q$. 

The radius of the sphere is fixed to $L_0(N)$, which depends on $N$ and was defined so that the mean Gaussian energy $\langle S_1\rangle/N$ is approximately equal to $\langle S_1\rangle/N\!\sim\!3/2$ as mentioned in the Introduction. We assumed the following values of $L_0(N)$: $L_0(N)\!=\!14$ for $N\!=\!1442$, $L_0(N)\!=\!18.6$ for $N\!=\!2562$, $L_0(N)\!=\!25.6$ for $N\!=\!4842$, and $L_0(N)\!=\!34$ for $N\!=\!8412$. Under those values of $L_0(N)$ for the radius, the relation $S_1/N\!=\!1.5$ is almost satisfied in the initial configurations for MC simulations. Note that the value of the radius $L_0(N)$, which satisfies $\langle S_1\rangle/N\!\sim\!3/2$, depends on the construction of the surface. Note also that the sphere with the radius $L_0(N)$ seems correspond to a real physical membrane at sufficiently low temperature or in the limit of $\alpha\to\infty$. 

The distance $2L(N)$ between two vertices is fixed to three different values such that
\begin{equation}
\label{Distance} 
L(N)=L_0(N), \; 1.5L_0(N), \; 2L_0(N)  
\end{equation} 
where $L_0(N)$ is a radius of an initial sphere constructed from the icosahedron as described above. Then, the distance between two fixed vertices on the surface with $L(N)=2L_0(N)$ is just $4L_0(N)$ for example. The boundary conditions are thus imposed on the model by these two fixed vertices of distance $2L(N)$. 

If it were not for the boundary condition defined in Eq. (\ref{Distance}), we have $\langle S_1\rangle/N\!=\!3/2$, which comes from the scale invariant property of $Z$. However, as we will see later, this relation is slightly broken due to the boundary condition.

\begin{figure}[hbt]
\centering
\includegraphics[width=9.5cm]{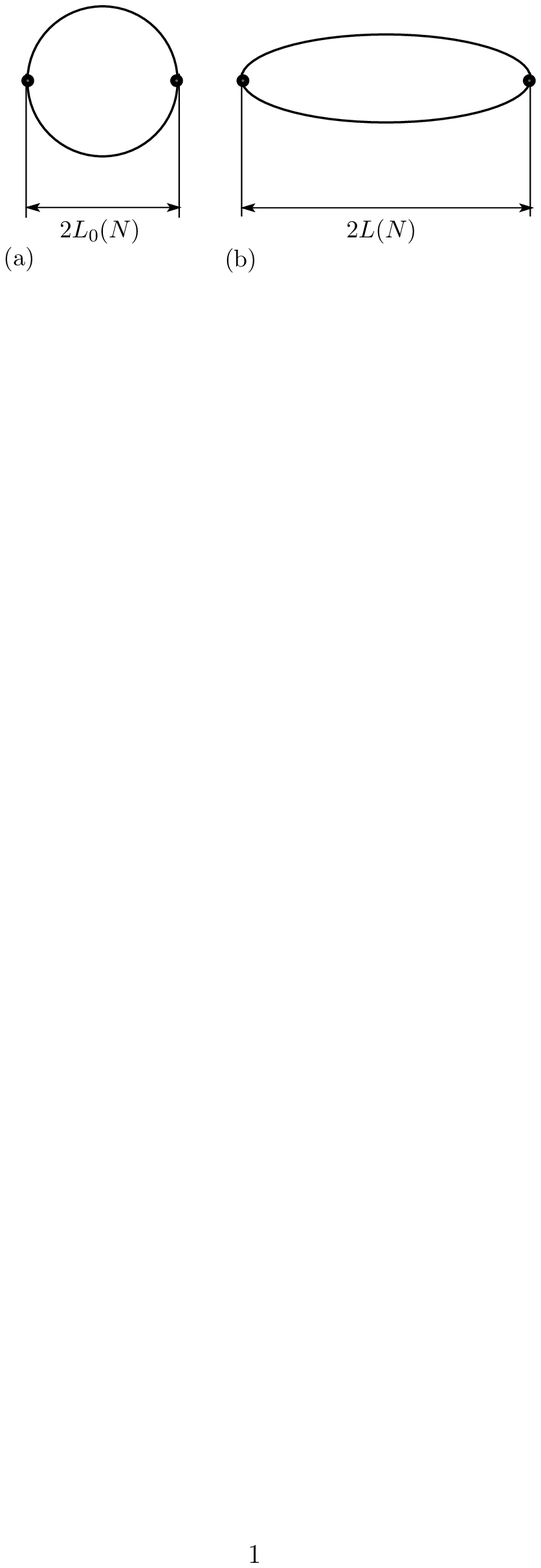}
\caption{Schematic drawing of (a) the diameter $2L_0(N)$ of a sphere which is the starting configuration for the MC simulations, and that of (b) the distance $2L(N)$ between two vertices of the expanded spherical surface, where $L(N)$ is fixed to three different values $L(N)=L_0(N)$, $L(N)=1.5L_0(N)$, and $L(N)=2L_0(N)$ in the simulations. Surfaces were drawn as simple as possible, since they are aimed only at showing $L_0(N)$ and $L(N)$. The symbols ($\bullet $) denote two fixed vertices on each surface. The surface in (b) is obtained by expanding the distance of the vertices from $2L_0(N)$ to $2L(N)$ during thermalization in MC.  }
\label{fig-1}
\end{figure}

Figure \ref{fig-1}(a) shows a schematic view of the diameter $2L_0(N)$ of a sphere, and Fig. \ref{fig-1}(b) shows that of the distance $2L(N)$ between the two vertices of the expanded spherical surface. Surfaces were drawn as simple as possible, since they are aimed only at showing $L_0(N)$ and  $L(N)$. 

The procedure for choosing and fixing the position of the boundary vertices is as follows: Firstly, we choose a pair of vertices which are on a straight line passing through the center of the triangulated sphere for the starting configuration. The canonical $x$-coordinate axis in ${\bf R}^3=\{(x,y,z)\vert x,y,z \in {\bf R}\}$ is chosen as the straight line in the surface. The distance between these two vertices on the axis is $2L_0(N)$ in the starting configuration. Secondly, the distance is expanded from $2L_0(N)$ to $2L(N)$ along the axis during the thermalization in MC, where $L(N)$ is given in Eq. (\ref{Distance}). The total number of the thermalization in MC is about $1\!\times\!10^7$ for expanding the distance. Some additional thermalization MCS are performed after the expansion, if necessary.

When the condition $L(N)\!=\!2L_0(N)$ was chosen, the distance $2L(N)$ is expanded from $2L_0(N)\!=\!14$ to $2L(N)\!=\!28$ during the thermalization for the $N\!=\!1442$ surface. The two fixed vertices of distance $2L_0(N)\!=\!14$ are shifted by distance $7\times 10^{-5}$ at every 100 MCS to the opposite direction with each other along the axis during the first $1\times 10^7$ thermalization MCS. Note also that the results of the simulation is completely independent of how the surface with the distance $2L(N)$ is constructed from the starting configuration with the distance $2L_0(N)$ between the boundary points. 

We comment on a relation between the deficit angle term $S_3$ and the integration measure $dX_i$ of the partition function in Eq. (\ref{Part-Func}). The integration measure $\prod_i dX_i$ can be replaced by  $\prod_i q_i^{\alpha} dX_i$, where $q_i$ is the co-ordination number of the vertex $i$ \cite{DAVID-NPB-1985}. This $\alpha$ is believed to be $2\alpha\!=\!3$. Then, it is possible to consider that $q_i^{\alpha}$ is the volume weight of the vertex $i$ in the measure $dX_i$. Hence, we can extend $2\alpha$ to non-integer numbers by assuming that the weight is arbitrarily chosen. Moreover, we can also extend $q_i$ to continuous numbers for the same reason. Hence, the weight $\prod_i q_i^{\alpha}$  can be replaced by $\prod_i \delta_i^{\alpha}\!=\!\exp ( \alpha \sum_i \log \delta_i )$. Including a constant weight $(2\pi)^\alpha$, we have $S_3$ in Eq. (\ref{Disc-Eneg}). 

It should also be noted that $S_3(\delta)$ in Eq. (\ref{Disc-Eneg}) can make the surface smooth not only in the model with the Gaussian term but also in the Nambu-Goto surface model within the class of tethered surfaces \cite{KOIB-PRE-2004-2}. Whereas the term $S_3(q)\!=\!-\sum_i \log q_i$ is constant on the tethered surfaces, and moreover the equivalent term $S_3(q)\!=\!\sum_i (q_i-6)^2$ plays no role in smoothing fluid surfaces \cite{KOIB-PRE-2003}. We therefore have discovered a significant difference for the role of $S_3(\delta)$ in Eq. (\ref{Disc-Eneg}) and that of the corresponding term $S_3(q)$ for smoothing the surface. 

The vertices $X$ are shifted so that $X^\prime \!=\! X\!+\!\delta X$, where $\delta X$ is chosen randomly in a small sphere. The new position $X^\prime$ is accepted with the probability ${\rm Min}[1,\exp(-\Delta S)]$, where $\Delta S\!=\! S({\rm new})\!-\!S({\rm old})$. We use two sequences of random numbers called Mersenne Twister \cite{Matsumoto-Nishimura-1998}; one for three dimensional random shift of $X$ and the other for Metropolis accept/reject. The radius of the small sphere for the shift $\delta X$ is chosen so that the rate of acceptance for $X$ is about $50\%$. We introduce the lower bound $1\times 10^7$ for the area of triangles. No lower bound is imposed on the bond length.

\section{Results}
Monte Carlo simulations were performed on the surfaces of size $N\!=\!1442$,  $N\!=\!2562$,  $N\!=\!4842$, and $N\!=\!8412$. The total number of sweeps of MC simulations were about $0.8\times 10^8\sim 1.2\times 10^8$ at transition point $\alpha_c$ for each surface. Relatively small numbers of sweeps were performed at non-transition points $\alpha\not\!=\! \alpha_c $. 

\begin{figure}[hbt]
\centering
\includegraphics[width=12.5cm]{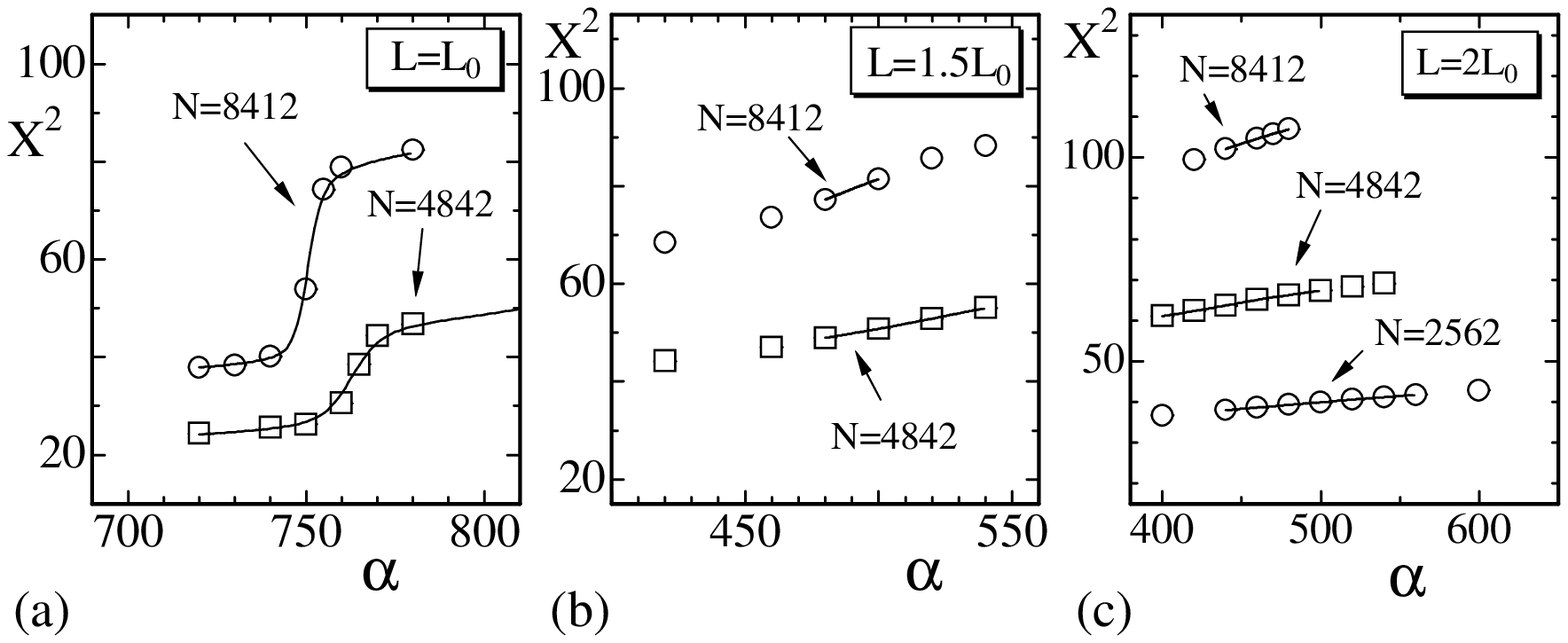}
\caption{$X^2$ vs. $\alpha$ obtained under the conditions (a) $L(N)\!=\!L_0(N)$, (b) $L(N)\!=\!1.5L_0(N)$, and (c) $L(N)\!=\!2L_0(N)$. Solid lines were obtained by multihistogram reweighting technique.}
\label{fig-2}
\end{figure}
First, we plot the mean square size $X^2$ in Figs. \ref{fig-2}(a)--\ref{fig-2}(c), where $X^2$ is defined by 
\begin{equation}
\label{X2}
X^2={1\over N} \sum_i \left(X_i-\bar X\right)^2, \quad \bar X={1\over N} \sum_i X_i,
\end{equation}
where $\bar X$ is the center of the surface. $X^2$ reflects the size of surfaces even on surfaces with the conditions given by Eq. (\ref{Distance}) for two fixed vertices. The boundary conditions for these data in Figs. \ref{fig-2}(a), \ref{fig-2}(b), and \ref{fig-2}(c) are given by $L(N)\!=\!L_0(N)$, $1.5L_0(N)$, and $2L_0(N)$, respectively. Solid lines drawn on the data were obtained by multihistogram reweighting technique \cite{Janke-histogram-2002}. The reweighting analysis was done by using all the simulation data in Fig. \ref{fig-2}(a), and it was done by using some of the data in Figs. \ref{fig-2}(b),\ref{fig-2}(c). 

Figure \ref{fig-2}(a) shows discontinuous changes of $X^2$ against $\alpha$ for $L(N)\!=\!L_0(N)$, whereas $X^2$ continuously changes in Figs. \ref{fig-2}(b),\ref{fig-2}(c) corresponding to  $L(N)\!=\!1.5L_0(N)$ and $L(N)\!=\!2L_0(N)$. The discontinuity of $X^2$ indicates that the model undergoes a first-order transition under the condition $L(N)\!=\!L_0(N)$.

\begin{figure}[hbt]
\centering
\includegraphics[width=12.5cm]{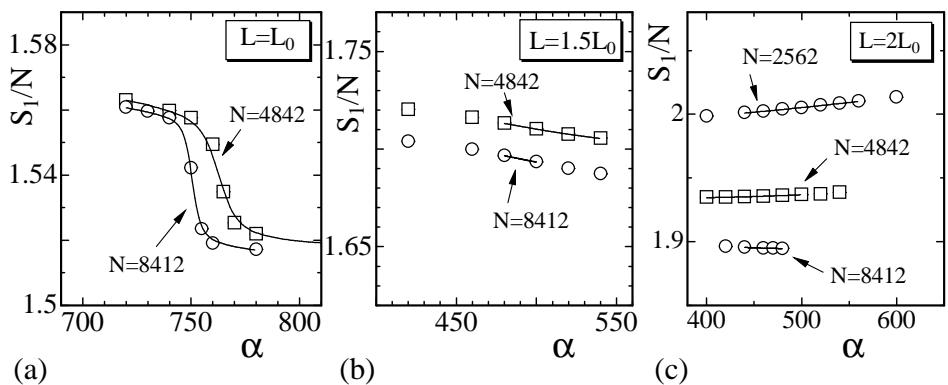}
\caption{$S_1/N$ vs. $\alpha$ obtained under the conditions (a) $L(N)\!=\!L_0(N)$, (b) $L(N)\!=\!1.5L_0(N)$, and (c) $L(N)\!=\!2L_0(N)$. Solid lines were obtained by multihistogram reweighting technique.}
\label{fig-3}
\end{figure}
We plot the Gaussian energy $S_1/N$ against $\alpha$ in Figs. \ref{fig-3}(a)--\ref{fig-3}(c). As mentioned in the previous section, $S_1/N$ should be equal to $3/2$ whenever no specific boundary conditions are imposed on the model. However, $S_1/N$ may deviate from $3/2$ in Fig. \ref{fig-3} because the surface is fixed with two vertices separated by a distance $2L(N)$. In fact, we see a discontinuous change of $S_1/N$ in Fig. \ref{fig-3}(a), which corresponds to the condition $L(N)\!=\!L_0(N)$. The discontinuity of $S_1/N$ indicates that the model undergoes a first-order transition under $L(N)\!=\!L_0(N)$.  We can also find that $S_1/N$ deviates from $3/2$ in Figs. \ref{fig-3}(b),\ref{fig-3}(c).

\begin{figure}[hbt]
\centering
\includegraphics[width=12.5cm]{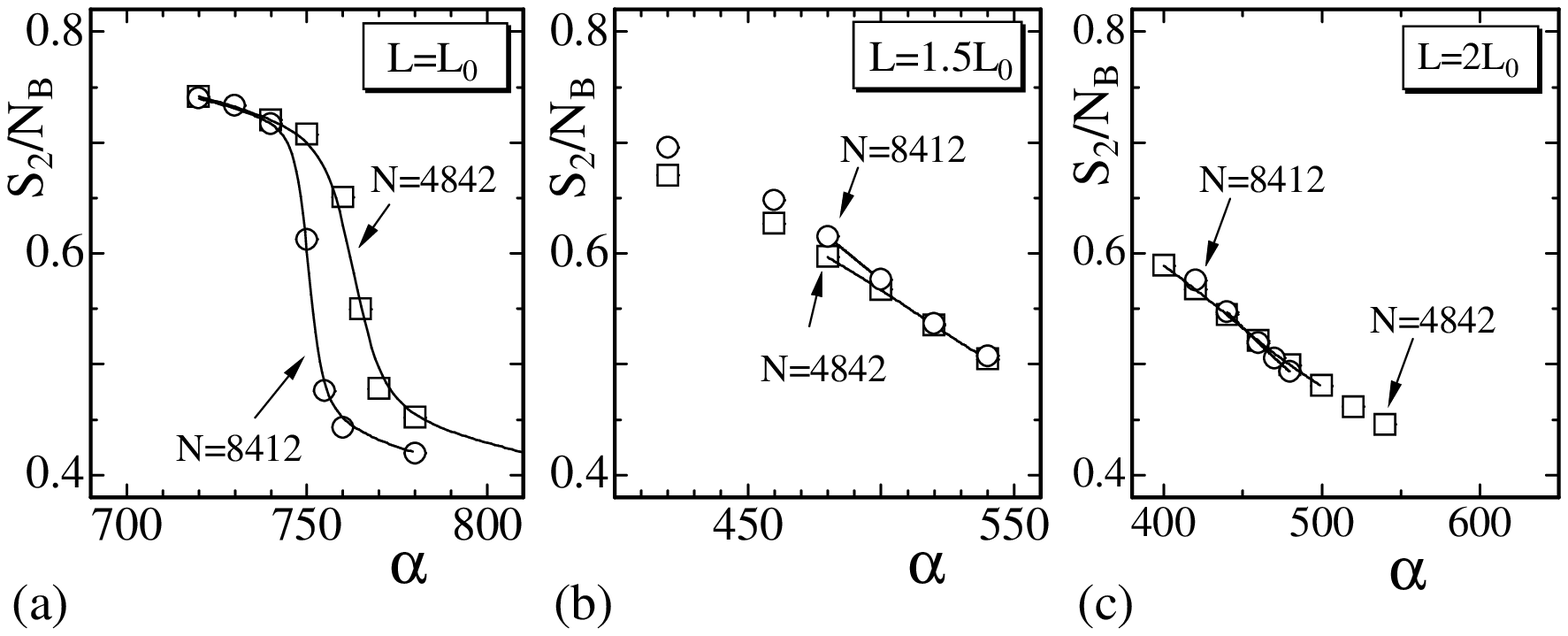}
\caption{The bending energy $S_2/N_B$ vs. $\alpha$ obtained under the conditions (a) $L(N)\!=\!L_0(N)$, (b) $L(N)\!=\!1.5L_0(N)$, and (c) $L(N)\!=\!2L_0(N)$. $S_2$ is not included in the Hamiltonian. $N_B$ is the total number of bonds. }
\label{fig-4}
\end{figure}
The bending energy $S_2/N_B$ is plotted in Figs. \ref{fig-4}(a)--\ref{fig-4}(c), where $N_B$ is the total number of bonds; $N_B\!=\!3N-6$. $S_2/N_B$ reflects the smoothness of the surface, although it is not included in the Hamiltonian. A phase transition can be called a first-order one if some physical quantity discontinuously changes. Therefore, the discontinuous change of $S_2/N_B$ in Fig. \ref{fig-4}(a) also supports the first-order transition of the model under the condition $L(N)\!=\!L_0(N)$.

\begin{figure}[hbt]
\centering
\includegraphics[width=12.5cm]{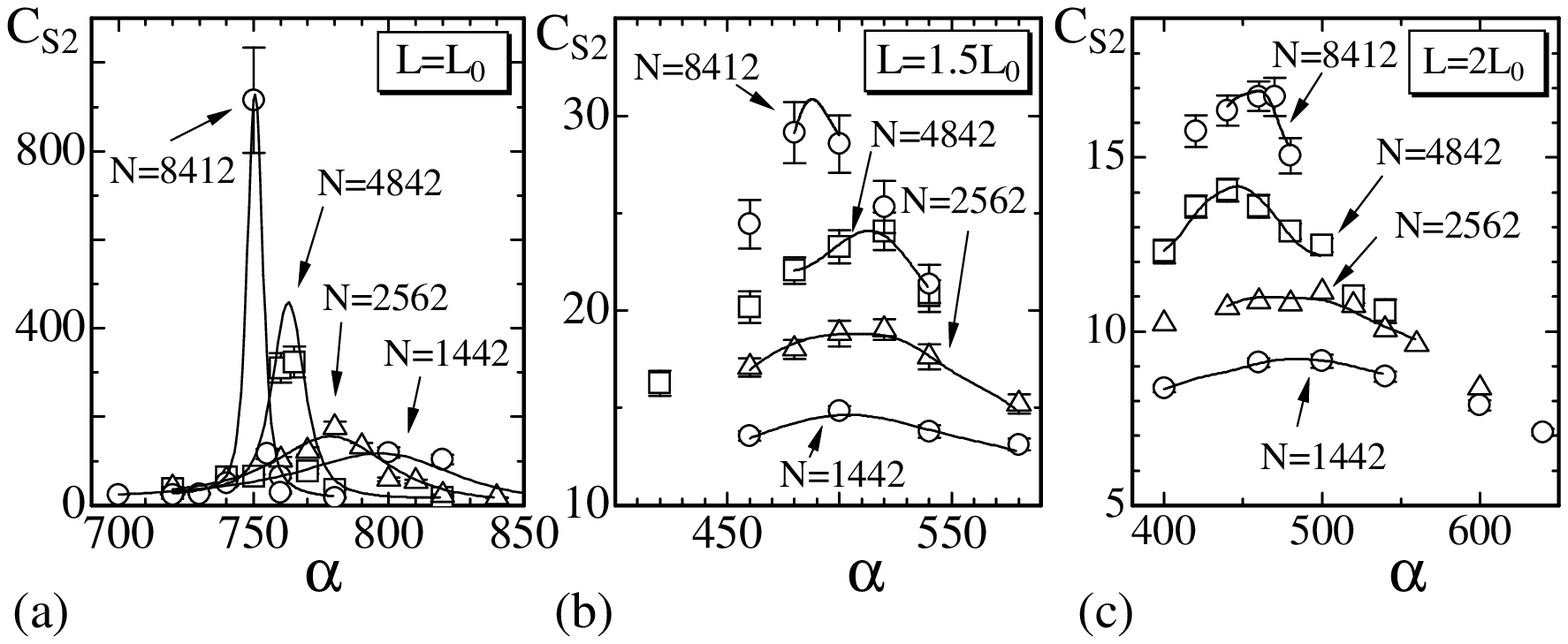}
\caption{The specific heat $C_{S_2}$ vs. $\alpha$ obtained under the conditions (a) $L(N)\!=\!L_0(N)$, (b) $L(N)\!=\!1.5L_0(N)$, and (c) $L(N)\!=\!2L_0(N)$.  }
\label{fig-5}
\end{figure}
The specific heat $C_{S_2}$ for the bending energy $S_2$ is defined by
\begin{equation}
\label{Spec-Heat}
C_{S_2} = {1\over N} \langle \; \left( S_2 - \langle S_2 \rangle\right)^2\rangle, 
\end{equation}
which is the variance of $S_2$. Note that the coefficient $b^2$, which is the squared bending rigidity, is excluded from the right-hand side of Eq. (\ref{Spec-Heat}). The reason for this is because the term $bS_2$ is not included in the Hamiltonian. However, we expect that an anomalous behavior of $C_{S_2}$, which is typical of the phase transition, is independent of whether the coefficient $b^2$ is included in $C_{S_2}$ or not.   

Figures \ref{fig-5}(a), \ref{fig-5}(b) and \ref{fig-5}(c) show $C_{S_2}$ against $\alpha$ obtained under the conditions $L(N)\!=\!L_0(N)$, $1.5L_0(N)$, and $2L_0(N)$, respectively. The error bars on the symbols in the figures are the statistical errors, which were obtained by the so-called the binning analysis. The solid lines were drawn by the multihistogram reweighting technique. We see the expected anomalous behavior of $C_{S_2}$ in Fig. \ref{fig-5}(a), which reflects a discontinuous transition. It is also clear from Figs. \ref{fig-5}(b) and \ref{fig-5}(c) that the transition is softened with the increasing distance $L(N)$; the peak values of $C_{S_2}$  become lower and lower when $L(N)$ increases.   

\begin{figure}[hbt]
\centering
\includegraphics[width=12.5cm]{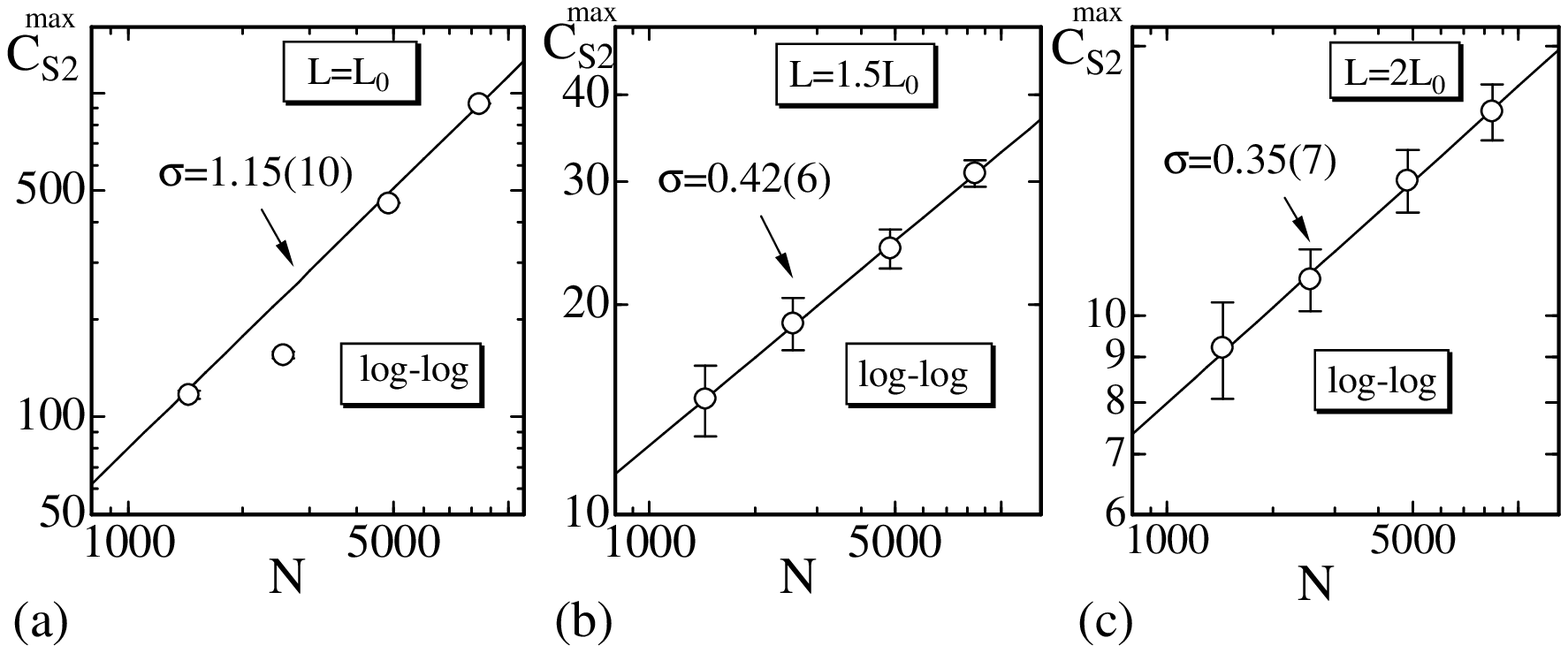}
\caption{The peak values $C_{S_2}^{\rm max}$ vs. $N$ obtained under the conditions (a) $L(N)\!=\!L_0(N)$ in a log-log scale, (b) $L(N)\!=\!1.5L_0(N)$ in a log-log scale, and (c) $L(N)\!=\!2L_0(N)$ in a linear-log scale.  $C_{S_2}^{\rm max}$ and the statistical errors were obtained by the multihistogram reweighting. The error bars denote the statistical errors.}
\label{fig-6}
\end{figure}
We obtained the peak value $C_{S_2}^{\rm max}$ of the specific heat by the multihistogram reweighting technique. The statistical errors were also obtained by that technique  and  shown in Figs. \ref{fig-6}(a)--\ref{fig-6}(c) with the error bars. The errors were almost invisible in  Fig. \ref{fig-6}(a), because they were relatively smaller than those shown in Figs. \ref{fig-6}(b) and \ref{fig-6}(c). The reason why the errors in Figs. \ref{fig-6}(b) and \ref{fig-6}(c) are relatively large seems that the number of data point in the simulations is not so large; overlapping of energy is slightly insufficient for the reweighting analysis in those cases.  

$C_{S_2}^{\rm max}$ is expected to scale according to 
\begin{equation}
\label{sigma-fitting}
C_{S_2}^{\rm max}  \sim N^\sigma,
\end{equation}
where $\sigma$ is a critical exponent of the transition. By fitting the data in Figs. \ref{fig-6}(a) and \ref{fig-6}(b) to Eq. (\ref{sigma-fitting}), we have 
\begin{eqnarray}
\label{sigma-result}
&&\sigma_1=1.15\pm 0.10,\quad [L(N)\!=\!L_0(N)],\nonumber \\
&&\sigma_2=0.42\pm 0.06,\quad [L(N)\!=\!1.5L_0(N)], \\
&&\sigma_3=0.35\pm 0.07,\quad [L(N)\!=\!2L_0(N)]\nonumber.
\end{eqnarray}
The value of $\sigma_1\!=\!1.15(10)$ indicates that the transition is of the first order, whereas $\sigma_2\!=\!0.42(6)$ and  $\sigma_3\!=\!0.35(7)$  indicate second-order transitions. 

The values $\sigma_2\!=\!0.42(6)$ and $\sigma_3\!=\!0.35(7)$ in Eq. (\ref{sigma-result}) can be compared to a known value $\sigma\!=\!0.58(10)$ of the model with extrinsic curvature reported in \cite{WHEATER-NPB1996}. However, we have no definite conclusion about whether the two models are in the same universality class, because $\sigma_2$ and $\sigma_3$ slightly deviate from $\sigma\!=\!0.58(10)$. Nevertheless, it is possible that two models are in the same class, because $\sigma$ of our model changes depending on the distance $L(N)$. It is also expected that the transition disappear when $L(N)$ further increases.  We should note that a possibility of a discontinuous change of $\sigma$ against $L(N)$ in the limit of $N\!\to\!\infty$ is not eliminated. 

\begin{figure}[hbt]
\centering
\includegraphics[width=12.5cm]{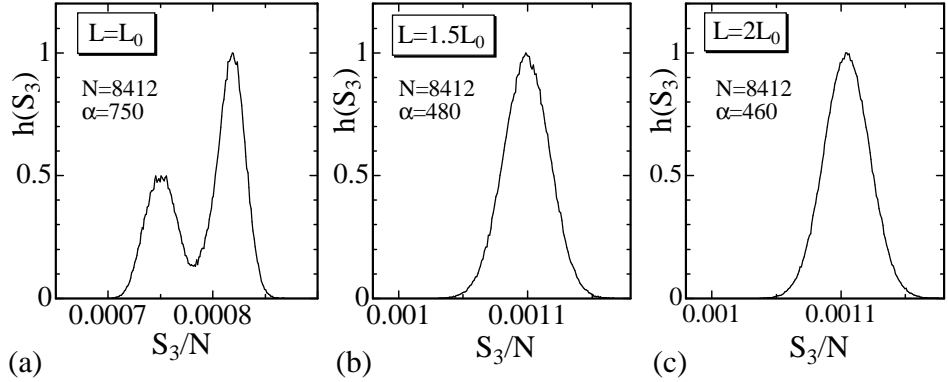}
\caption{The histogram $h(S_3)$ of $S_3/N$ obtained at the transition point of the $N\!=\!8412$ surface under the conditions (a) $L(N)\!=\!L_0(N)$, (b) $L(N)\!=\!1.5L_0(N)$, and (c) $L(N)\!=\!2L_0(N)$. A double peak structure of $h(S_3)$ in (a) indicates a first-order transition. }
\label{fig-7}
\end{figure}
The first-order transition should be reflected in $S_3/N$ such that $S_3/N$ discontinuously changes at the transition point. In order to see this, we plot the distribution (or histogram) of the intrinsic curvature energy $S_3/N$ in Figs. \ref{fig-7}(a)--\ref{fig-7}(c). We find that $h(S_3)$ has a double peak structure in Fig. \ref{fig-7}(a) under the condition $L(N)\!=\!L_0(N)$. This clearly shows that $S_3/N$ is discontinuous and that the model undergoes a first-order transition at $\alpha\!=\!750$ on the $N\!=\!8412$ surface. On the contrary, $S_3/N$ smoothly changes in Figs. \ref{fig-7}(b),\ref{fig-7}(c) under the conditions $L(N)\!=\!1.5L_0(N)$ and $L(N)\!=\!2L_0(N)$.

We can also see a discontinuous change of $S_3/N$ at $\alpha\!=\!750$ in a plot of $S_3/N$ against $\alpha$, which is not depicted here. However, the discontinuity of $S_3/N$ is not so clear. Therefore, we can hardly clarify the order of the transition of the model under $L(N)\!=\!L_0(N)$ without the histogram $h(S_3)$ shown in Fig. \ref{fig-7}(a). We understand that the transition is softened by the conditions of Eq. (\ref{Distance}) although it remains in the first-order transition only under the condition $L(N)\!=\!L_0(N)$.

\begin{figure}[hbt]
\vspace{0.5cm}
\unitlength 0.1in
\begin{picture}(0,0)(  0,0)
\put(9,33){\makebox(0,0){(a) {$L\!=\!L_0,\;\alpha\!=\!750({\rm smooth})$}}}%
\put(32,33){\makebox(0,0){(b) {$L\!=\!L_0,\;\alpha\!=\!750({\rm crumpled})$}}}%
\put(32,-2){\makebox(0,0){(d) The surface section of (b)}}%
\put(9,-2){\makebox(0,0){(c) The surface section of (a)}}%
\end{picture}%
\vspace{0.5cm}
\centering
\includegraphics[width=10.5cm]{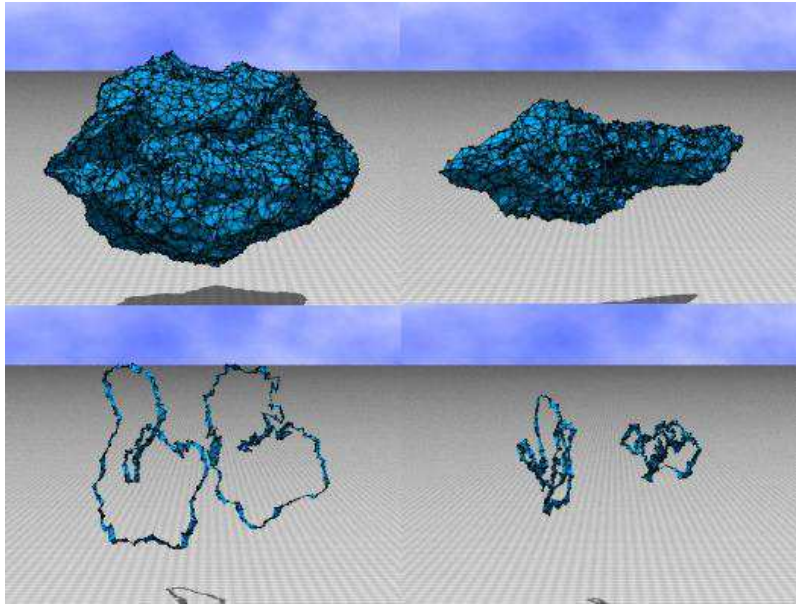}
\caption{Snapshots of surfaces in (a) the smooth phase and (b) the crumpled phase at the transition point $\alpha\!=\!750$ of the $N\!=\!8412$ surface under $L(N)\!=\!L_0(N)$,  (c) surface sections in (a), and (d) surface sections in (b).    }
\label{fig-8}
\end{figure}
Snapshots of surfaces in the smooth phase and the crumpled phase are respectively shown in Figs. \ref{fig-8}(a) and \ref{fig-8}(b), which are obtained at the transition point $\alpha\!=\!750$ of the $N\!=\!8412$ surface under $L(N)\!=\!L_0(N)$. The surface sections are shown in  Figs. \ref{fig-8}(c), \ref{fig-8}(d).

\begin{figure}[hbt]
\vspace{0.5cm}
\unitlength 0.1in
\begin{picture}(0,0)(  0,0)
\put(6.5,33){\makebox(0,0){(a) {$L\!=\!1.5L_0,\;\alpha\!=\!480$}}}%
\put(29,33){\makebox(0,0){(b) {$L\!=\!2L_0,\;\alpha\!=\!460$}}}%
\put(32,-2){\makebox(0,0){(d) The surface section of (b)}}%
\put(9,-2){\makebox(0,0){(c) The surface section of (a)}}%
\end{picture}%
\vspace{0.5cm}
\centering
\includegraphics[width=10.5cm]{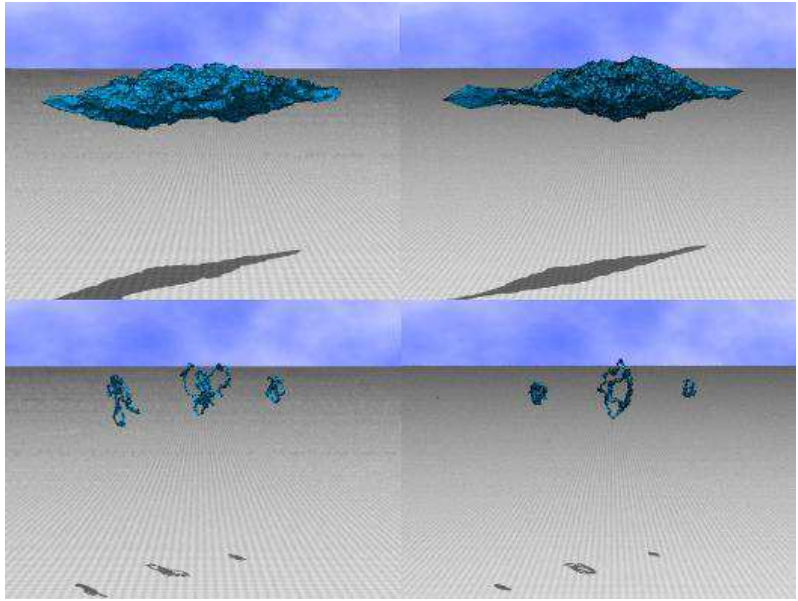}
\caption{Snapshots of surfaces (a) at the transition point $\alpha\!=\!480$ of the $N\!=\!8412$ surface under $L(N)\!=\!1.5L_0(N)$, (b) at the transition point $\alpha\!=\!460$ of the $N\!=\!8412$ surface under $L(N)\!=\!2L_0(N)$, (c) surface sections in (a), and (d) surface sections in (b). }
\label{fig-9}
\end{figure}
 Figures \ref{fig-9}(a) and \ref{fig-9}(b) are snapshots of surfaces obtained  at the transition point $\alpha\!=\!480$ of the $N\!=\!8412$ surface under $L(N)\!=\!1.5L_0(N)$ and at the transition point $\alpha\!=\!460$ of the $N\!=\!8412$ surface under $L(N)\!=\!2L_0(N)$, respectively. The surface sections in Figs. \ref{fig-9}(a) and  \ref{fig-9}(b) are shown in Figs. \ref{fig-9}(c) and  \ref{fig-9}(d), respectively.  

We understand from the snapshots in Figs. \ref{fig-8} and \ref{fig-9} that the surfaces become oblong if the distance $L(N)$ is increased. Therefore, it is possible to extract the string tension of surfaces from MC data if $L(N)$ is sufficiently large, although the model is defined on a fixed connectivity surface. However, the string tension of the model will be continuous at the transition point because the transition becomes a second-order or  higher-order one when $L(N)$ is sufficiently large. This implies that the string tension does not play the role of an order parameter in the crumpling transition of a fixed connectivity surface model. However, it is very interesting that the strength of the crumpling transition changes depending on the boundary condition.

\section{Summary and conclusions}
We have investigated the phase structure of a surface model with an intrinsic curvature under a boundary condition such that two fixed vertices are separated by a distance $2L(N)$. The model was known as the one that undergoes a first-order crumpling transition without the boundary condition \cite{KOIB-EPJB-2004,KOIB-PLA-2005-1,KOIB-PLA-2006-1}. This paper aimed to show how boundary conditions influence the phase transition, and we performed extensive MC simulations on the spherical tethered surfaces up to a size $N\!=\!8412$.

The distance $L(N)$ was chosen to be three different values for each $N$; $L(N)\!=\!L_0(N)$, $L(N)\!=\!1.5L_0(N)$, and $L(N)\!=\!2L_0(N)$, where $L_0(N)$ depends on $N$ and is the radius of the initial sphere at the start of the MC simulations.  We used the following values of $L_0(N)$: $L_0(N)\!=\!14$ for $N\!=\!1442$, $L_0(N)\!=\!18.6$ for $N\!=\!2562$, $L_0(N)\!=\!25.6$ for $N\!=\!4842$, and $L_0(N)\!=\!34$ for $N\!=\!8412$, so that the relation $S_1/N\!=\!1.5$ is almost satisfied at the starts. 

We have found that the transition is softened as the distance $L(N)$ is increased. The order of the transition remains in the first-order under $L(N)\!=\!L_0(N)$, and it changes to the second-order when the distance is increased to $L(N)\!=\!1.5L_0(N)$ and  $L(N)\!=\!2L_0(N)$, where the peak $C_{S_2}^{\rm max}$ of the specific heat for the bending energy scales according to $C_{S_2}^{\rm max}\sim N^\sigma, \sigma<1$ at the transition point. The first-order transition at $L(N)\!=\!L_0(N)$ was confirmed with a double peak structure in the histogram $h(S_3)$ of the intrinsic curvature energy $S_3/N$, which is included in the Hamiltonian.

The result is in sharp contrast to that of a fluid surface model with extrinsic curvature, where the crumpling transition is strengthened when the distance between two fixed vertices is increased under a specific condition \cite{KOIB-PLA-2004,KOIB-EPJB-2005} at a sufficiently large $L(N)$. In fact, the phase transition of the tethered surface model in this paper is softened at sufficiently large $L(N)$ as stated above. Moreover, the phase transition is expected to disappear if $L(N)$ was further increased, where the surface becomes sufficiently oblong. Therefore, we consider that string tension does not play the role of an order parameter in the crumpling transition if the model is a fixed connectivity one at least. Nevertheless, the fact that the strength of the crumpling transition changes depending on the boundary condition is very interesting, because the boundary condition seems accessible in real physical membranes. If the transition was observed in a biological membrane that is governed by the intrinsic curvature, the strength of the transition can be handled by fixing and expanding the surface.   

It is interesting to study the phase structure of the model on dynamically triangulated fluid surfaces under the same boundary condition, where the surface is expected to be oblong and linear if the distance $L(N)$ is increased to a sufficiently large size. For future experimental studies on the crumpling transition in biological membranes, further numerical studies on the surface models will provide more detailed and helpful information about the influence of boundary conditions on the transition. 

\section*{Acknowledgment}
This work is supported in part by a Grant-in-Aid of Scientific Research, No. 15560160. The author (H.K) acknowledges the staff of the Technical Support Center in Ibaraki College of Technology for their support in computer analysis. 



\end{document}